\documentclass[12pt]{iopart}

\usepackage{iopams}
\usepackage{graphicx}
\usepackage{color}

\begin{document}

\title{Direct Imaging of Slow, Stored, and Stationary EIT Polaritons}

\author{Geoff T Campbell$^{1,3}$, Young-Wook Cho$^1$, Jian Su$^1$, Jesse Everett$^1$, Nicholas Robins$^2$, Ping Koy Lam$^1$, Ben Buchler$^1$}
\address{$^1$Centre for Quantum Computation and Communication Technology,\\Department of Quantum Science, The Australian National University,
Canberra, ACT 2601, Australia}

\address{$^2$Department of Quantum Science, The Australian National University, Canberra, ACT 2601, Australia}

\ead{$^3$geoff.campbell@anu.edu.au}

\begin{abstract}
Stationary and slow light effects are of great interest for quantum information applications. Using laser-cooled Rb87 atoms we have performed side imaging of our atomic ensemble under slow and stationary light conditions, which allows direct comparison with numerical models. The polaritions were generated using electromagnetically induced transparency (EIT), with stationary light generated using counter-propagating control fields. By controlling the power ratio of the two control fields we show fine control of the group velocity of the stationary light.  We also compare the dynamics of stationary light using monochromatic and bichromatic control fields.  Our results show negligible difference between the two situations, in contrast to previous work in EIT based systems.
\end{abstract}

\maketitle

\newcommand{\bra}[1]{\langle #1 \vert}
\newcommand{\ket}[1]{\vert #1 \rangle}
\newcommand{\inner}[2]{\langle #1 \vert #2 \rangle}

\section*{Introduction}

Electromagnetically induced transparency (EIT) \cite{Fleischhauer:RevEIT:2005,Marangos1998_EITReview,Hau:1999} is a technique that enables fine control of the propagation of light fields. It is ordinarily achieved using a \textit{probe} beam that copropagates with a \textit{control} beam through an atomic ensemble.  In the absence of the control light, the probe is absorbed into the atomic ensemble. The addition of control light induces transparency for the probe, which can then pass through the ensemble with, potentially, very little absorption. Modulation of the control field allows the slowing and even storage of light within the atomic medium. Accordingly, EIT has been used to demonstrate the slowing, storage and retrieval of quantum states of light \cite{PhysRevA.65.022314,Akamatsu:2004bu,Chaneliere:2005fu,Eisaman:2005hp,Akamatsu:2007fy,Hetet:2008uy,Appel:2008,Kimble_nature_sph-QM1,Honda:2008p4680,Zhou:2012fw} and has been proposed to have uses in sensing applications \cite{fleischhauer2000quantum,leonhardt2000ultrahigh} where slow light can be advantageous.  It has also been extended to demonstrate a wide range of coherent control techniques to manipulate light. One interesting example is that of stationary light, where two counter-propagating control fields are used to stop the propagation of an EIT polariton even though a portion of the excitation remains optical \cite{Andre2002,bajcsy2003,Andre2005,moiseev2006quantum,Moiseev2007,lin2009}. The optical part can be used to generate a nonlinear interaction, necessary for  optical quantum gates \cite{Harris1999,Chen:2006cf,Chen2012,Feizpour:2015bt}.

In most stationary light experiments to date, only the optical field at the output end of the ensemble has been detected and compared to theoretical predictions. It is possible, however, to use side-imaging of a cold atomic ensemble to directly observe the dynamics of EIT polaritons. The technique has been applied in Bose Einstein condensates \cite{Ginsberg:2007ih,Zhang:2009fi} and warm atomic vapours \cite{wilson2016slow} to observe the dynamics of EIT slow light. Recently, we have also used side-imaging to directly observe the dynamics of polaritons in an off-resonant stationary light scheme \cite{Everett:2016eb} that was performed in a cold atomic cloud.

Here, we use the side-imaging technique to observe the propagation of EIT-based polaritons in a laser-cooled atomic ensemble under various conditions. The results are in good agreement with simulations of a simple three-level model and illustrate a range of effects such as slow light, storage, backward retrieval, and stationary light. The experiments provide strong evidence in support of the existence of stationary light under EIT conditions. We further demonstrate that the group velocity of propagating light can be precisely controlled by changing the ratio of forward and backward propagating control fields.

One aspect of this system, that is the subject of ongoing investigations, is the impact of atomic motion on the dynamics of EIT-based stationary light.  When the counter-propagating control fields form a standing wave there is potential for rapid decay of the stationary light, but this depends on the temperature of the atoms \cite{lin2009,Nikoghosyan2009b,Wu2010d,Peters2012a,Iakoupov2016a,Blatt:2016jq}.  In our experiments we investigate the impact of the standing wave control field and compare this situation to a bichromatic control field without a standing wave.  Our experiments show no significant difference between these two cases, in contrast to previous work.  We speculate on the reasons for this and offer suggestions for further work that may resolve this issue.

\subsection*{Slow-Light with Counter-Propagating Control Fields}

The dynamics of EIT slow-light can be understood via the Maxwell-Bloch equations of motion. In the pure-state and paraxial approximations, these equations can be written \cite{Gorshkov:2007:PRA:76} as
\begin{eqnarray}
\partial_t P = -\Gamma P + i g \sqrt{N} \mathcal{E} + i \Omega S \label{eq:MB_Simple:1} \\
\partial_t S = -\gamma S + i \Omega^* P \label{eq:MB_Simple:2}\\
(\partial_t + c \partial_z) \mathcal{E} = i g \sqrt{N}  P \label{eq:MB_Simple:3}.
\end{eqnarray}
Referring to the atomic level scheme shown in Fig.~\ref{fig:Setup}b, $P(z,t)$ is the envelope of the excited state coherence $|3\rangle \leftrightarrow |2\rangle$, $S(z,t)$ is the envelope of the spin coherence $|1\rangle \leftrightarrow |3\rangle$ and $\mathcal{E}(z,t)$ is the envelope of the probe field. Decay rates $\Gamma$ and $\gamma$ are imposed for $P$ and $S$, respectively, $g$ is the probe field coupling rate, $N$ is the number of atoms and $\Omega$ is the control field Rabi frequency. If the envelope $\mathcal{E}(z,t)$ varies slowly enough, $P \approx 0$ and $\mathcal{E} \approx -\Omega/(g\sqrt{N})S$. The light and the atomic spin coherence then propagate together as a polariton defined as $\psi = (\mathcal{E}\sin\theta - S\cos\theta)$ where $\tan\theta = \Omega/(g\sqrt{N})$ \cite{PhysRevA.65.022314}. The equation of motion for the polariton is $\left(\partial_t + \sin^2\theta c \partial_z + \cos^2\theta\gamma \right) S = 0$
which, in a transformed a set of coordinates $\tau = t - z/c$, $\xi = z/L$, can be written
\begin{equation}
  \left(\partial_\tau + \frac{c}{L} \tan^2\theta \partial_\xi + \gamma \right) \psi = 0
\end{equation}
where $L$ is the length of the ensemble.

To determine the dynamics of slow light with counter-propagating control fields, we treat the forward and backward traveling components of $P$ independently by defining $P = P_+e^{ikz} + P_-e^{-ikz}$, where $k$ is the wavenumber of the excited state coherence \cite{moiseev2006quantum,lin2009}. These are coupled to the atomic ground state and meta-stable state through $\mathcal{E}_\pm$ and $\Omega_\pm$ respectively, as illustrated in Fig.~\ref{fig:Setup}(b). Standing waves formed by the counter-propagating control and probe fields will create higher spatial frequencies in $S$ of the form $S = S_0 + \sum{^\infty_{n=1}S_{n+} e^{i2nkz} + S_{n-} e^{-i2nkz}}$. We expect a rapid decay of the higher spatial frequencies of the coherence due to atomic motion and truncate to $n=1$.

Because $\mathcal{E}_{\pm}$ are counter-propagating, there is no coordinate transformation that will remove the time derivative from the propagation equations for both the forward and backward traveling waves. We can, however, omit the time derivatives by assuming that the speed of light is sufficiently large that $L/c$, which on the order of $10^{-10}$ s, is much shorter than any other timescale in the system dynamics. We then define a spatial coordinate
\begin{equation}
  \xi(z) = \int_0^z dz' \eta(z')/N \label{eq:xi}
\end{equation}
that is scaled by the relative optical density along the ensemble, $\eta(z)$. Introducing the optical depth $d = g^2 N L/(\Gamma c)$ and scaling $\mathcal{E}$ by a dimensionless factor of $\sqrt{c/(\Gamma L)}$ allows us to write the Maxwell-Bloch equations in a compact form
\begin{eqnarray}
P_{\pm} = i\sqrt{d} \mathcal{E}_{\pm} + i (\Omega_{\pm}/\Gamma) S_0 \label{eq:MB_CP:1} + i (\Omega_\mp/\Gamma) S_{\pm}\\
\partial_t S_0 = -\gamma S_0 + i \Omega_+^* P_+ + i \Omega_-^* P_- \label{eq:MB_CP:2}\\
\partial_t S_\pm = -\tilde\gamma S_\pm + i \Omega_\mp^* P_\pm \label{eq:MB_CP:3}\\
\partial_\xi \mathcal{E}_{\pm} = i \sqrt{d} P_{\pm} \label{eq:MB_CP:4}
\end{eqnarray}
where we have made the adiabatic approximation $\Gamma P_\pm \gg \partial_t P_\pm$. The higher spatial frequencies are assumed to decay at a rate $\tilde\gamma$ as a result of averaging due to atomic motion. These higher spatial frequencies have been shown to result in additional diffusion, decay and pulse splitting of the stationary polariton \cite{Nikoghosyan2009b,Wu2010d,Peters2012a,Iakoupov2016a} for sufficiently cold atoms.

To find a compact analytic solution, we assume that the higher spatial frequencies $S_\pm$ dissipate sufficiently quickly to be negligible, although we include them for numerical solutions \cite{lin2009}. The equations can then be solved in the Fourier domain of the normalised spatial coordinate $X(\xi,t) = \int d\kappa e^{-i\kappa\xi} X(\kappa,t)$ \cite{zimmer2008dark}. Combining equations (\ref{eq:MB_CP:1}) and (\ref{eq:MB_CP:4}) and expanding to the first order in $\kappa/d$ we obtain
\begin{equation}
  \mathcal{E}_\pm \simeq -\frac{\Omega_\pm}{\sqrt{d}\Gamma} \left( 1 \pm i \kappa/d \right)S.
\end{equation}
The quantity $\kappa/d$ is the spatial variation of the envelope of $S$ relative to the absorption depth of the ensemble. Substituting this into Eq.~\ref{eq:MB_CP:2} and transforming out of the Fourier domain yields an equation of motion
\begin{equation}
  \left[ \partial_t + \Gamma \tan^2\theta\left(\cos{2\phi}\partial_\xi - \frac{1}{d}\partial_{\xi\xi}\right) + \gamma \right] S = 0 \label{advection}
\end{equation}
where
\begin{equation}
  \tan^2\theta \equiv \frac{\vert\Omega\vert^2}{d\Gamma^2}; \quad \tan^2\phi \equiv \frac{\vert\Omega_-\vert^2}{\vert\Omega_+\vert^2}; \quad \vert\Omega\vert^2 \equiv \vert\Omega_+\vert^2 + \vert\Omega_-\vert^2.
\end{equation}
This is a shape-preserving advection equation with a velocity $v = \Gamma \tan^2\theta\cos{2\phi}$ and a diffusion term $(\Gamma/d) \tan^2\theta \partial_{\xi\xi} S$. The diffusion is due to finite optical depth and arises in the equation of motion due to taking the more relaxed adiabatic approximation $\partial_t P \ll \Gamma P_\pm$ instead of $P_\pm \approx 0$. In the limit of large optical depth or slow pulses, $k/d \ll 1$, the diffusion is negligible and a polariton $\psi = \sin\theta\left(\mathcal{E}_{+}\cos\phi+\mathcal{E}_{-}\sin\phi\right) - S \cos\theta$ can be defined for the system. The mixing angles $\theta$ and $\phi$ are governed by the total control field power and the ratio between the power of the forward and backward control fields, respectively. Either mixing angle can be used to reduce the polariton velocity to zero and $\phi$ can be used to reverse the direction of propagation. 

\section*{Experimental methods}

\begin{figure}[b!]
\centering
\includegraphics[width=0.85\linewidth]{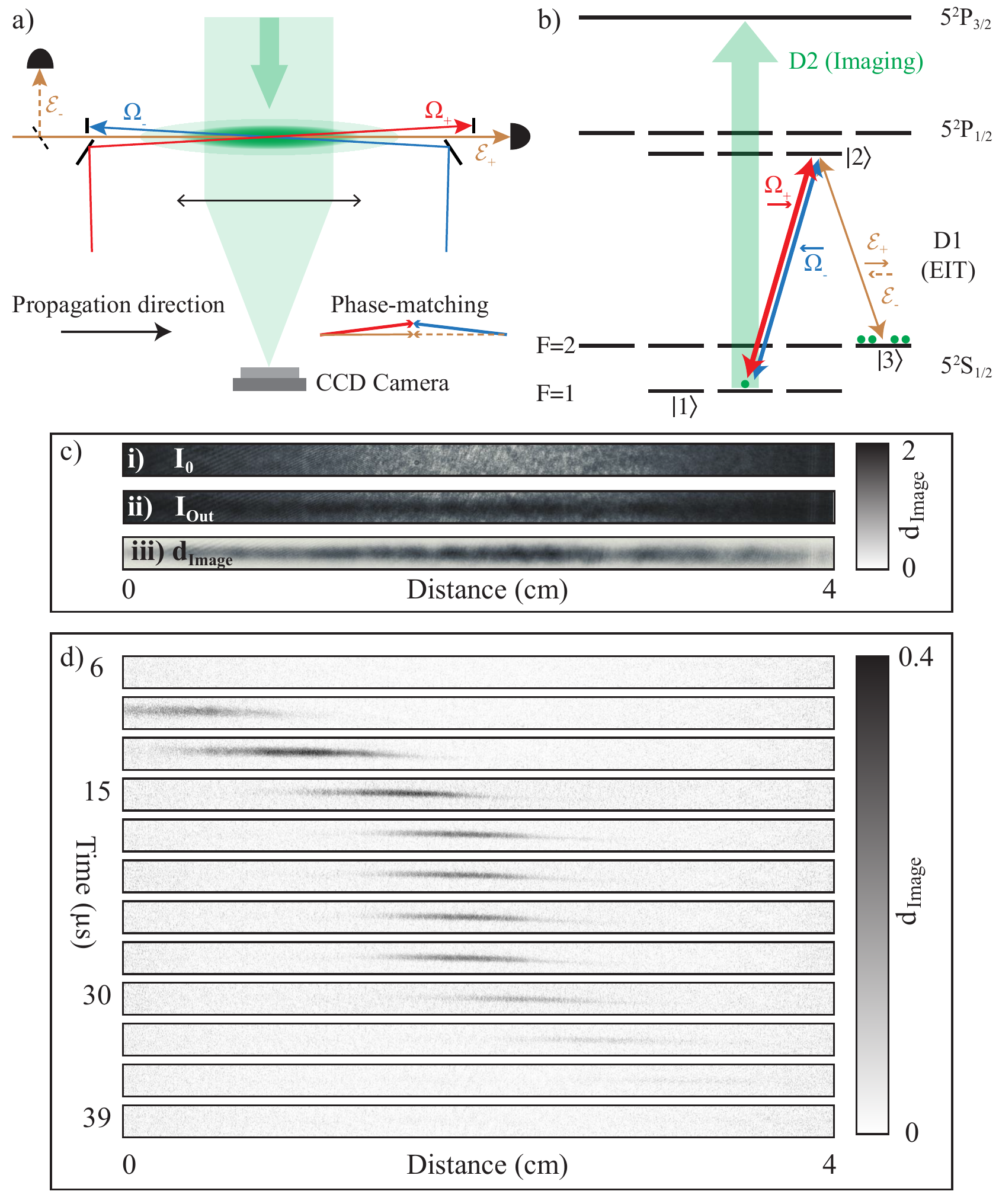}
\caption{(a) Schematic of the experimental layout. (b) Energy level diagram showing the EIT tripod and imaging beam. (c) Characterization of the atom distribution in the MOT showing the imaging beam intensity $I_0$, transverse absorption from the MOT $I_z$ and calculated transverse optical depth $d_{image}$. (d) Absorption images taken every 3 $\mu$s of an EIT polariton traveling through the ensemble. Slow light, storage and release can be seen.}
\label{fig:Setup}
\end{figure}

We experimentally analysed the dynamics of slow-light by sending a probe pulse into an elongated cloud of cold atoms \cite{Cho:16,Sparkes:2013njp} that was uniformly illuminated by a bright control field. The magnitude of the spin coherence as the pulse propagated along the length of the cloud could then be observed by absorption imaging from the side of the cloud. We first prepared the atoms in an elongated magneto-optical trap (MOT), which provided an ensemble of atoms with a temperature of approximately 100 $\mu$K and an amplitude optical depth of $d = 190$ along the trap axis. The atom cloud was approximately 4 cm long with a cross-section of 200 $\mu$m. The probe field was aligned along the cloud axis and  focused with a beam diameter of 110 $\mu$m. The control fields were collimated to a larger diameter to ensure uniform coverage of the atomic ensemble.  

Figure \ref{fig:Setup} shows the layout of the experiment (a) and the atomic transitions used for EIT and imaging (b). Phase-matching between the forward and backwards EIT processes was achieved by placing the control fields on the transition with the shorter wavelength and selecting an appropriate angle relative to the probe. The output forward- and backward-travelling probe fields were detected using photo-diodes after the control fields were removed by spatial filtering with pinholes (not shown). The absorption imaging was done by illuminating the ensemble from the side with light tuned to the D$_2$ transition. The shadow of the atoms was imaged onto a CCD camera using a large aperture lens.

The magnitude of $\vert S \vert$ is proportional to the number of atoms, which in turn is proportional to the optical depth as seen by the imaging beam.  This can be found according to $d_{image} = -\log{(I_{out}/I_{0})}/2$ where $I_0$ is the intensity of the imaging beam in the absence of any atoms.  Figure \ref{fig:Setup}(c) shows images of $I_0$, $I_{out}$, and $d_{image}$ for the entire atom cloud. This was obtained by optically pumping the ensemble into $|1\rangle$ which is the transition used for imaging.

For a slow light experiment, the cloud is initially prepared in the $|3\rangle$ state and a probe pulse is sent into the cloud while the forward control field is on to build some coherence between the $|1\rangle$ and $|3\rangle$. The CCD camera is exposed for 300 $\mu$s, spanning the entire duration of a slow-light or storage experiment, but the imaging beam is gated to illuminate the ensemble for only 1 $\mu$s. This allows a stroboscopic measurement of the location of the spin coherence as it travels through the cloud. Repeating the experiment and shifting the exposure time in 1~$\mu s$ intervals allowed composition of a space-time image of the spinwave propagation. Figure \ref{fig:Setup}(d) shows samples of the images taken of pulse propagation during slow-light, storage, and release.

To map $\vert S \vert$ into the normalised spatial coordinate $\xi$, the images can be binned along the propagation dimension according to the optical density of the atom cloud, as in Eq.~\ref{eq:xi}. This binning reduced the length of each image from 1384 to 200 bins, each of which contains a roughly equal number of atoms. The transverse region of the image that contains the pulse was then integrated to obtain a one-dimensional array proportional to $\vert S \vert$.  In the following results we will present data showing scaled values of $|S|$ as a function of time and $\xi$ to allow comparison with numerical models.

\section*{Results}

\subsection*{Storage and Retrieval}

\begin{figure}[b!]
\centering
\includegraphics[width=\linewidth]{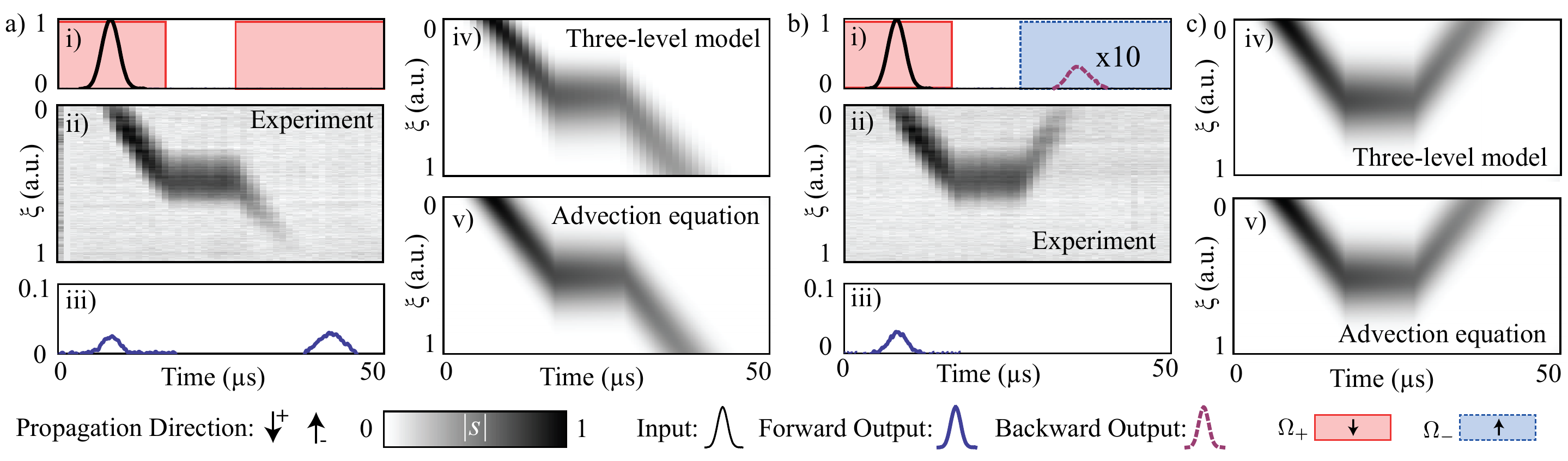}
\caption{Propagation of an EIT polariton as it experiences slow-light, storage, and either forward retrieval (a) or backward retrieval (b). The experimentally measured propagation (ii) is compared to the results of a numerical simulation (Eqs.~\ref{eq:MB_CP:1}-\ref{eq:MB_CP:4}) (iv) and to the solution of a shape-preserving advection equation (Eq.~\ref{advection}) (v). The optical field at the input of the ensemble (i) and output (iii) is measured by photo-detectors. Some transmission is visible on the output photo-detector; this is a spurious frequency component of the probe field that is far-detuned from resonance and is an artifact from how we generate the probe field. The values of $|S|$ are scaled to the maximum value of $|S|$.}
\label{fig:FSL}
\end{figure}

Figure \ref{fig:FSL}(a) shows the propagation of the coherence $\vert S \vert$ in the normalised coordinate as a probe pulse enters the ensemble under the conditions of EIT slow-light. As is shown in the timing diagram (i), a pulse enters the ensemble while the control field is on and can be seen in the imaging data propagating slowly through the ensemble (ii). The propagation is stopped by turning off the control field, and is then resumed by restoring the control field. The optical pulse emerges from the ensemble and is recorded on a photo-detector (iii). The shape-preserving nature of the slow-light pulse propagation can be seen to be in good agreement with numerical simulations of Eqs.~\ref{eq:MB_CP:1}-\ref{eq:MB_CP:4}  (iv), and the simple advection Eq.~\ref{advection} (v). All of the parameters used in the simulations correspond to those used in the experiment and were measured independently.

Figure \ref{fig:FSL}(b) shows backward retrieval of a coherence using a backward-propagating control field. The recalled optical pulse measured on the photo-detector that records the counter-propagating probe field is shown on the timing diagram (i). In this case, the observed pulse propagation through the ensemble (ii) is in good agreement with both a three-level model (iv) and the advection equation (v).

\subsection*{Stationary Light}

\begin{figure}
\centering
\includegraphics[width=\linewidth]{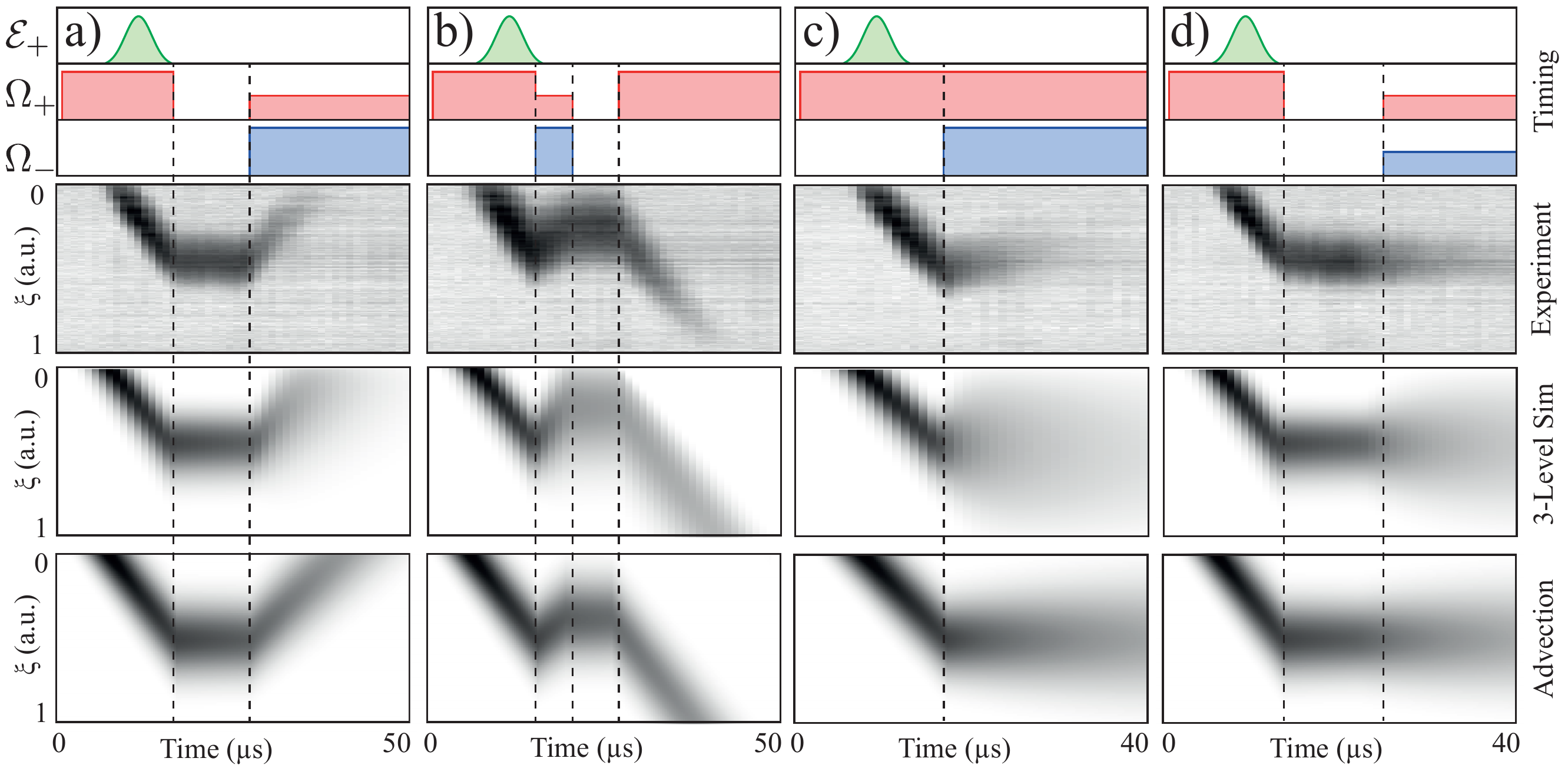}
\caption{Experimental polariton propagation compared to simulations and solutions of an advection equation for various control field timings and amplitudes. (a) Forward slow light (FSL), storage (S) and retrieval using quasi-stationary light with imbalanced control fields (QSL). (b) A sequence of FSL, QSL, S, FSL. (c) FSL followed by stationary light. (d) FSL, S, stationary light with half-intensity control fields.  The values of $|S|$ are scaled to the maximum value of $|S|$.}
\label{fig:Results_Matrix}
\end{figure}

In addition to using either the forward or backward control fields, both control fields can be used simultaneously to modify the propagation of the polariton by changing the mixing angle $\phi$ in Eq.~\ref{advection}. Figure \ref{fig:Results_Matrix} shows a number of experiments demonstrating different slow-light effects with both control fields. For each experiment, the observed propagation data is shown along with the solutions to numerical simulations of Eqs.~\ref{eq:MB_CP:1}-\ref{eq:MB_CP:4}, and to the corresponding advection Eq.~\ref{advection}. Column (a) shows a reduced backward propagation velocity due to the addition of a forward control field with an amplitude that is half that of the backward control field. Column (b) shows a complex sequence of forward slow light, backward propagation with some forward control as well, stored light and then forward recall. The sequence shows that the use of counter-propagating control fields can be used to controllably push the coherence in either direction within the ensemble.

In columns (c,d), stationary light is demonstrated by illuminating the ensemble with both control fields at equal amplitude simultaneously, once the polariton has propagated to the centre of the cloud. In (c), the stationary light is formed directly from slow light by turning on the counterpropagating control field. In (d) The polariton is stopped by turning off the control field, and stationary light is formed by turning on both control fields at half of the initial amplitude. In both cases, the polariton is held nearly stationary while both control fields are on. From this, and from the agreement between the observed propagation dynamics and those that are predicted, we infer that a stationary optical field is present in the ensemble.

The diffusion of the polariton arises from limited optical depth, the standing wave pattern formed by the control fields, and thermal motion of the atoms in the cloud. The temperature of the atoms in our system was measured to be 100~$\mu$K in a previous experiment \cite{Cho:16}, giving a mean atomic velocity of 10 cm/s. This is slow enough that we can neglect diffusion due to atomic motion. The solution of the advection equation, Eq.~\ref{advection}, includes only the effect of limited optical depth while the numerical solutions of Eqns.~\ref{eq:MB_CP:1}-\ref{eq:MB_CP:4} also takes into account the standing wave. The decay rate of the higher spatial frequencies that we use in the simulation is estimated from the atomic thermal motion according to $\tilde\gamma = 4\pi\sqrt{k_B T/m}/\lambda = 2\pi\times 0.25$ MHz. We note that diffusion is actually reduced with increasing temperature because atomic motion becomes significant compared to the length scale of the standing wave but not compared to the length scale of the polariton. This eliminates the diffusion term resulting from the standing wave but diffusion that arises directly from thermal motion remains negligible. Our results, however, show less diffusion than would be expected for the measured temperature. 

\begin{figure}
\centering
\includegraphics[width=0.7\linewidth]{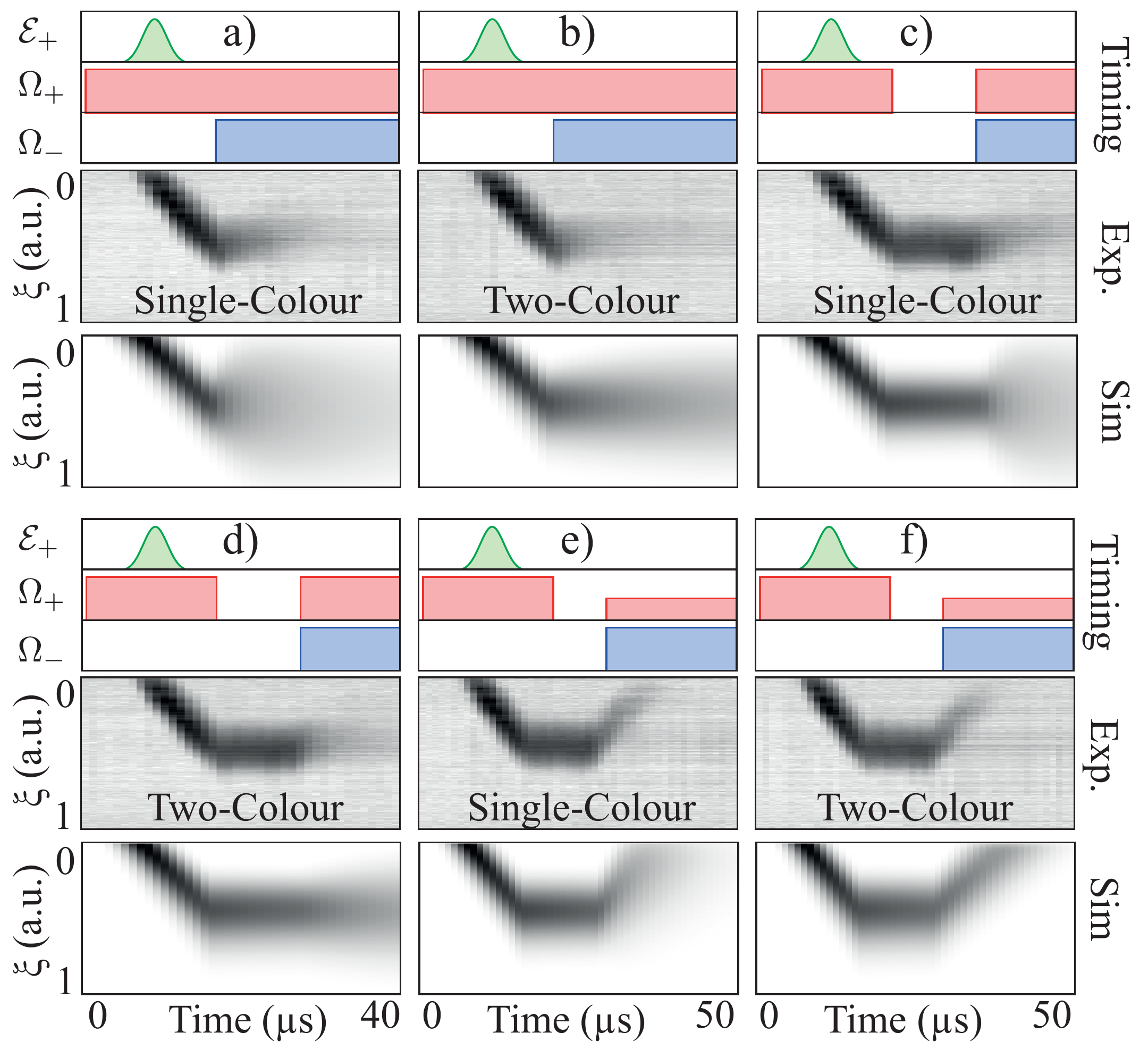}
\caption{A comparison of stationary light using control fields with equal frequencies (a,c,e) and with frequencies that are symmetrically detuned from resonance by 4 MHz (b,d,f). The stationary light is formed directly from slow-light (a,b) and from stopped light (c,d). Numerical simulations predict reduced dispersion when the control fields are detuned, however, a rapid decay of the polariton in the experimental case obscures the effect. Quasi-stationary light is observed (e,f) with imbalanced control field powers. Again, differences between the single- and two-color cases are not resolvable in the experiment.   The values of $|S|$ are scaled to the maximum value of $|S|$.}
\label{fig:Detuning_Comparison}
\end{figure}

To further investigate the diffusion we are able to manipulate the effective decay rate for the standing wave terms by introducing a frequency difference between the forward and backward propagating components \cite{lin2009}. In this case, the interference between counter-propagating control fields forms a travelling wave that averages out the fine spatial structure. Running the experiment in this regime may allow one to distinguish the diffusion of the stationary light due to finite optical depth and diffusion due to the standing wave of the control field. We therefore performed stationary light experiments with the control fields symmetrically detuned from the excited state transition by $\pm4$~MHz, for an effective decay rate of $\tilde\gamma_{eff} = \tilde\gamma + 2\pi\times8$~MHz.  We expect that this should quickly eliminate the diffusion due to the standing wave, leading to lower diffusion overall.

Figure \ref{fig:Detuning_Comparison} compares the dynamics of the single color stationary light with that of the two-color stationary light. We test the dynamics for stationary light from slow light (a,b), stationary light from stopped light (c,d), and slow propagation with unequal forward and backward control fields (e,f). The increased diffusion for the single-colour case is noticeable in the simulations, however, no significant difference can be seen in the experimental data.  This indicates that we are in a regime where the diffusion is dominated by the limited optical depth and the standing wave does not play a role.  One possible reason for this is that our temperature measurements are insensitive to atomic motion along the optical axis and may underestimate the effect of motional averaging if the velocity distribution of the atoms is anisotropic. In fact, we expect a larger atomic velocity along the optical axis because the optical pumping to prepare the atoms in the $|1\rangle$ state is performed by an axially aligned beam.

\subsection*{Reflection}

\begin{figure}[b!]
\centering
\includegraphics[width=1 \linewidth]{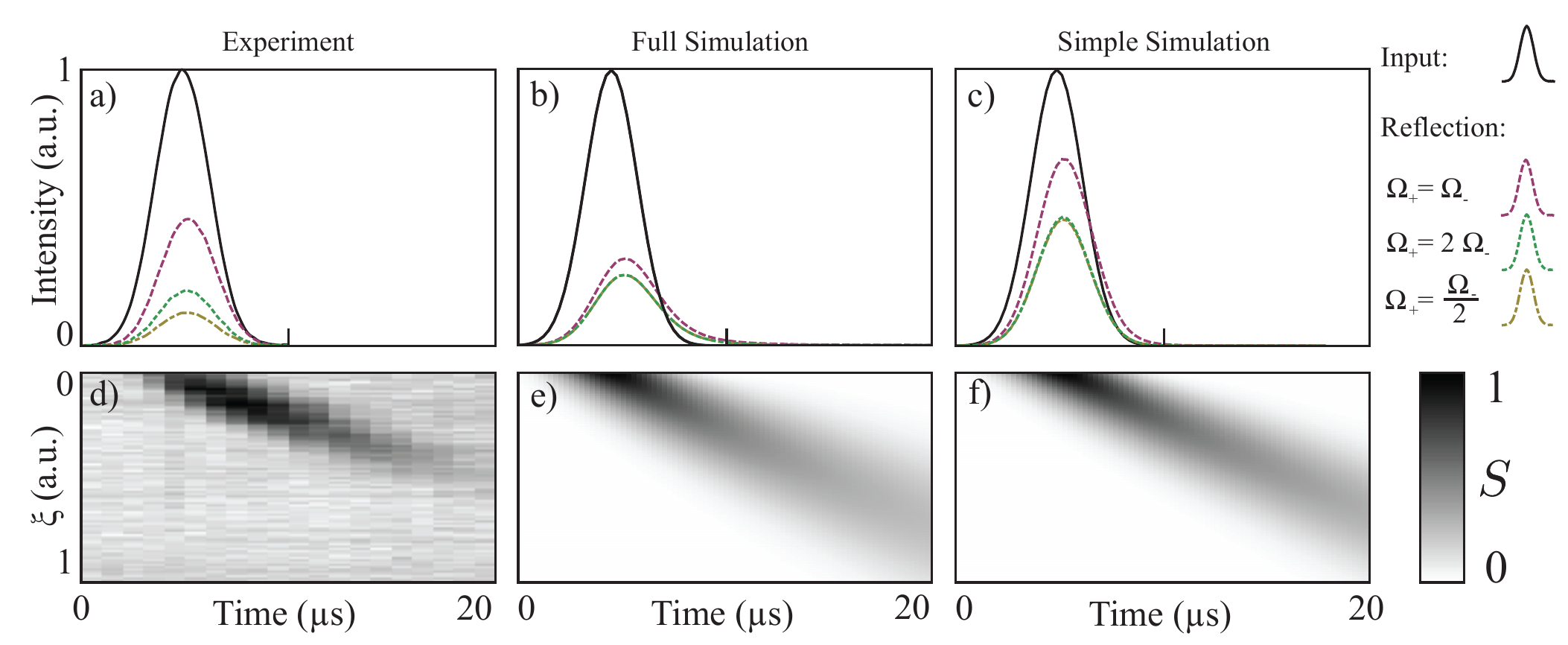}
\caption{Reflection off of the ensemble is observed for equal and imbalanced control field powers. Experimental data showing the power reflected from the atomic ensemble is shown in (a) and theoretical simulations with (b) and without (c) the standing wave terms. The bottom row shows data for the case of $\Omega_+=2\Omega_-$.  The experimental data (d) and models (e,f) demonstrate that a polariton propagates some way into the ensemble before being reemitted in the backward direction.  The values of $|S|$ are scaled to the maximum value of $|S|$.}
\label{fig:Reflection}
\end{figure}

All results presented thus far were obtained by first allowing the polariton to propagate into the ensemble as a slow-light polariton with only the forward control field before turning on the backward control to create stationary light. If both control fields are present when the probe pulse is incident on the ensemble, reflection can be observed. Figure \ref{fig:Reflection} (top-row) shows the observed (a) and simulated (b,c) reflection from the ensemble when it is illuminated with equal or unequal forward and backward control fields. The two simulations show the results that include the standing wave terms (b) and that neglect them (c). For the case where the forward control field is stronger than the backward control field, a polariton can be observed propagating into the ensemble (bottom-row). For the cases where the forward control field intensity is equal or less than that of the backward control field, no polarition can be seen entering the ensemble in the imaging data (not shown), although the simulations show a small region spin coherence at the start of the ensemble (not shown).


\subsection*{Group velocity control}

\begin{figure}[h!]
\centering
\includegraphics[width=0.8\linewidth]{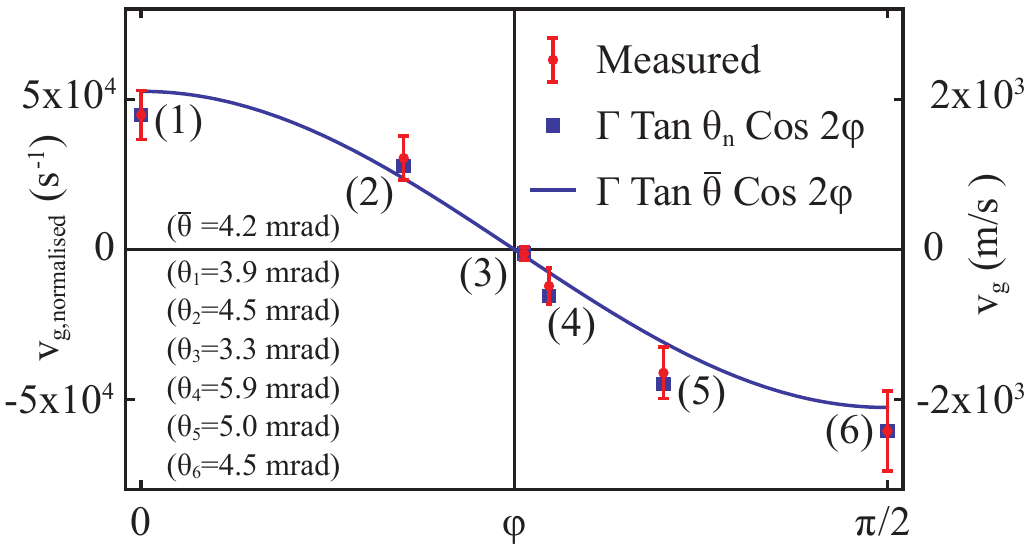}
\caption{Measured group velocities (red dots) for different control field powers compared to the expected values (blue squares). The mixing angle $\theta$ differed slightly between measurements and the values for each measurement are shown in the lower left quadrant. The solid curve shows the expected behavior if the total control field power had been held constant across all the measurements.}
\label{fig:GroupVel}
\end{figure}

The group velocity as a function of the mixing angles $\theta$ and $\phi$ can be directly determined from the imaging data. Figure \ref{fig:GroupVel} shows the measured group velocities along with those calculated for some of the combinations of control field powers used for the data presented in figs.~\ref{fig:FSL}-\ref{fig:Reflection}. The measured group velocities are compared to the expected values for each point based on the measured control field powers. There are slight differences in the values for control field powers used here and those used in the numerical simulations. Differences in the transverse size of the control fields resulted in a different calibration between the measured control power and the control Rabi frequency for the forward and backward controls. The numerical simulations were run with the intended ratio between forward and backward fields while the data in Fig.~\ref{fig:GroupVel} uses a calibration that is based on the measured group velocities of forward and backward slow light.

\section*{Discussion}

\begin{figure}
\centering
\includegraphics[width=\linewidth]{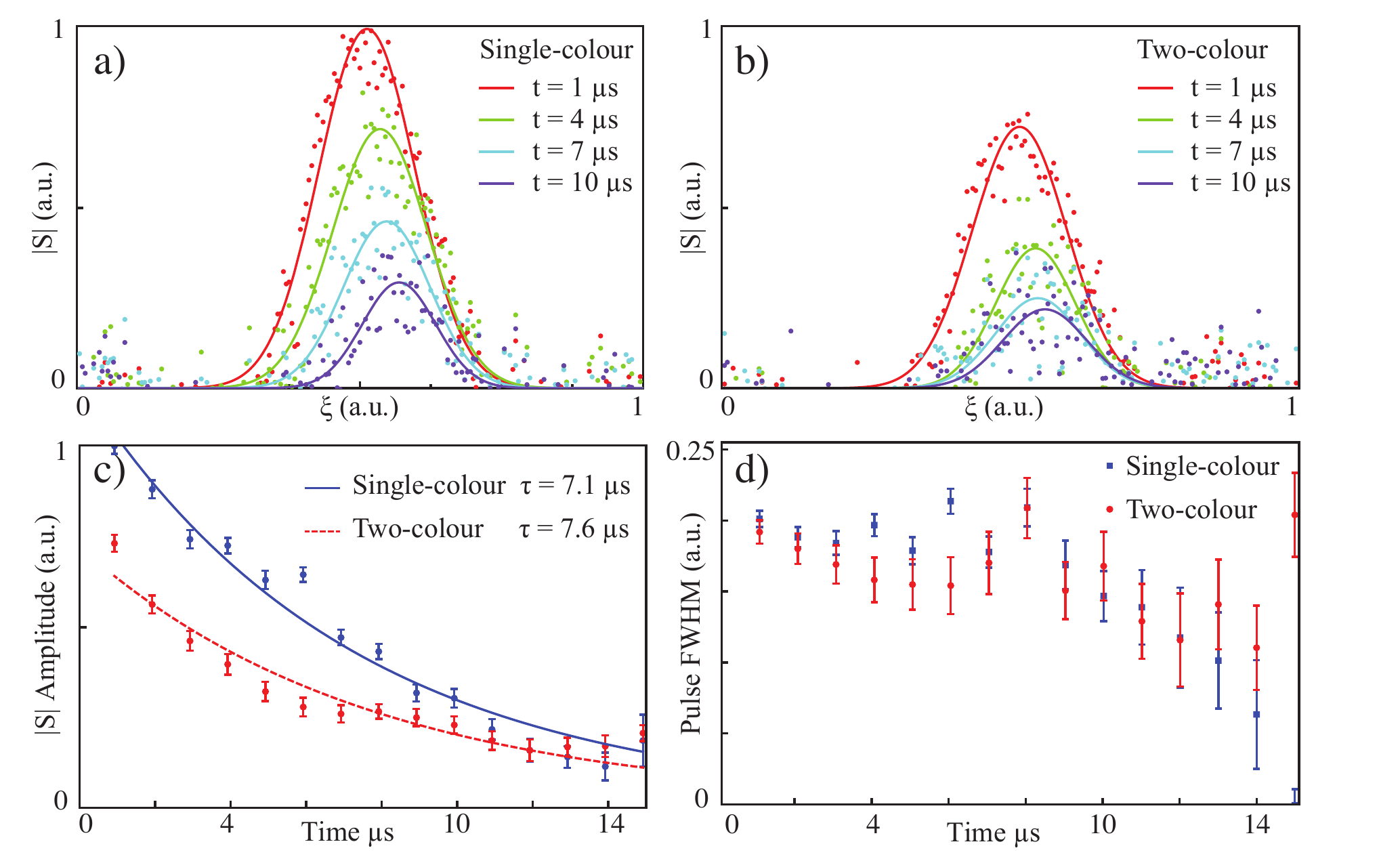}
\caption{a) Gaussian fits to spatial profile of the spinwave for single-colour stationary light.  b) Gaussian fits to spatial profile of the spinwave for two-colour stationary light. c) Decay of the spinwave amplitude for one- and two-colour stationary light. d) Spatial width of the spinwave as the stationary light decays for the one- and two-colour data.}
\label{fig:Dispersion}
\end{figure}

While our results are in generally good agreement with the theoretical models, there is stronger attenuation and less dispersion of the stationary polariton than predicted. To quantify both the attenuation and dispersion for stationary light, we fit Gaussian envelopes to spatial profiles of the spinwaves under stationary light conditions as shown in Fig.~\ref{fig:Dispersion} (a) for single-color stationary light and (b) for two-color stationary light.   These cross sections are made from the data presented earlier in Fig.~\ref{fig:Detuning_Comparison} (a,b).  Plotting the decay of the spinwave amplitude and the Gaussian full-width-half-maximum as a function of time we arrive at  Figs.~\ref{fig:Dispersion} (c) and (d) respectively. Figure~\ref{fig:Dispersion}(c)  shows that the time-constants for the decay of the polaritons are $7.1\pm1$~$\mu$s for the single-color case and $7.6\pm1.5$~$\mu$s for the two-color case respectively. Figure~\ref{fig:Dispersion}(d) shows that no diffusion of the pulses is apparent, although the pulse width fits have a large uncertainty due to noise in the images.  From these images we conclude that there is no observable difference between the monochromatic and bichromatic controls fields in our experiment.  This is in contrast to other observations and theoretical predictions \cite{lin2009,Nikoghosyan2009b,Wu2010d,Peters2012a,Iakoupov2016a,Blatt:2016jq} which predict a large diffusion due to the standing wave terms in the coherence.  As noted above, a possible cause of the discrepancy may be some net longitudinal motion of the ensemble.  The temperature measurement is insensitive to any motion of the cloud that is common to all atoms in the ensemble. Such motion, however, would still average out the high spatial frequencies associated with the standing wave of the counter-propagating control fields.

The observed decay time, shown in Fig.~\ref{fig:Dispersion} is significantly shorter than the estimated decoherence rate $\gamma \approx 500$ Hz for our ensemble. A possible cause of this attenuation is atomic population present in the $\ket{F=1, m_F = \lbrace -1,0 \rbrace}$ states. Transitions from these states will absorb the probe field since there is no corresponding control field to provide electromagnetically induced transparency. This residual Beer's law absorption can be easily included in the equations of motion, however, the initial slow light propagation seen in figures \ref{fig:FSL} and \ref{fig:Results_Matrix} is consistent with negligible additional absorption.

Another possible cause of the extra loss is that the background magnetic field, which is tuned to be zero at the time that the pulse enters the medium, changes over the course of the experimental run and leads to a misalignment of the quantization axis.  Further work is required to quantify the changes in the magnetic field over the duration of the experiment to determine how large this effect may be.

In terms of applications, EIT based stationary light appears to be less favourable than the self-stabilising stationary light scheme we have described in \cite{Everett:2016eb}, which was implemented with the same physical experimental setup. In the EIT regime the lifetime of the stationary light appears to be shorter, even without the expected impact of the standing wave terms.  While better control of the magnetic field environment may improve the lifetime, the sensitivity of the scheme to magnetic fields is still likely to be a disadvantage. 

\section*{Conclusion}

We have applied the technique of side absorption imaging to visualize the dynamics of stationary and non-stationary electromagnetically induced transparency polaritons when driven by counter-propagating control fields. Our results demonstrate that EIT stationary light can be modelled with a simple equation of motion. We have also shown how tuning the power ratio of the counterpropagating control fields allows fine control of the group velocity of the stationary light. Absent from the results are signatures that arise from high spatial frequencies due to standing wave control field. Further cooling our ensemble may reveal the modification of dynamics that is expected from the standing wave terms.

\section*{Acknowledgments}

We thank J.~R. Ott, A.~S. S{\o}rensen and their team for very helpful discussions on the role of higher-order spatial frequencies in the atomic coherence. Our work was funded by the Australian Research Council (ARC) (CE110001027, FL150100019).

\section*{References}
\bibliographystyle{unsrt}
\bibliography{EIT_SL_Bib}

\end{document}